\documentclass[twocolumn,showpacs,prb,aps]{revtex4}
\usepackage{amssymb}
\usepackage{graphicx}

\begin{document}

\title{Collective dynamics in molten potassium: an Inelastic X-ray Scattering study. }
\author{A.~Monaco$^1$, T.~Scopigno$^2$, P.~Benassi$^1$, A.~Giugni$^1$, G.~Monaco$^3$, M.~Nardone$^1$, G.~Ruocco$^2$, M.~Sampoli$^4$}
\date{\today}

\affiliation{
    $^{1}$Dipartimento di Fisica and INFM, Universit\'a di L'Aquila, I-67010, L'Aquila, Italy. \\
    $^{2}$Dipartimento di Fisica and INFM, Universit\'a di Roma ``La Sapienza'', I-00185, Roma, Italy.\\
    $^{3}$European Synchrotron Radiation Facility, B.P. 220 F-38043 Grenoble, Cedex France.\\
    $^{4}$Dipartimento di Energetica and INFM, Universit\'a di Firenze, I-50019, Firenze, Italy.
    }

\begin{abstract}
The high frequency collective dynamics of molten potassium has
been investigated by inelastic x-ray scattering, disclosing an
energy/momentum transfer region unreachable by previous neutron
scattering experiments (INS). We find that a two-step relaxation
scenario, similar to that found in other liquid metals, applies to
liquid potassium. In particular, we show how the sound velocity
determined by INS experiments, exceeding the hydrodynamic value by
$\approx 30 \%$, is the higher limit of a speed up, located in the
momentum region $1<Q<3$ nm$^{-1}$, which marks the departure from
the isothermal value. We point out how this phenomenology is the
consequence of a microscopic relaxation process that, in turn, can
be traced back to the presence of ''instantaneous'' disorder,
rather than to the crossover from a liquid to solid-like response.
\end{abstract}


\maketitle

\section{introduction}

Alkali metals have been known since long time to exhibit
remarkably pronounced inelastic features (Brillouin peaks) in
their density fluctuations frequency spectra, even for wavevector
($Q$) well beyond the region of validity of simple hydrodynamics.
For this reason, in the last thirty years, the microscopic
dynamics of such systems has been widely investigated both
experimentally, by means of Inelastic Neutron Scattering (INS),
\cite{cop_rb,mor_na,bod_cs,ver_li,nov_k,yulm_cs,cab_k,bov_k} and
numerically
\cite{rahman_sim,baluc1_sim,baluc2_sim,can_lit,kamb_sim,shim1_sim,foley_sim,scop_presim}
with the aim of clarifying in detail the mechanisms underlying the
atomic motions at a microscopic level.

More recently, the possibility of studying experimentally the
microscopic dynamics has been significantly increased by the
advent of Inelastic X-ray Scattering (IXS) spectroscopy
\cite{BURKEL,mas_strum}. Thanks to this technique, new insight
into the collective properties of alkali metals have been provided
\cite{sinn,pil_na,scop_epl,scop_prlli,scop_prbumk,scop_prena},
allowing to draw a much clearer picture of the high frequency
dynamics, so that a deep comprehension of the ultimate nature of
the high frequency excitations in these systems is gradually
emerging. In particular, thanks to development of approaches based
on the memory function formalism, \cite{BY,BALUCANI} a framework
has been developed which accounts for the relaxation spectra of
the density fluctuations in a large wavevector region
\cite{scop_jpc}. This framework applies also to non-alkali metals
\cite{scop_preal,scop_prlga}, hydrogen bonded liquids
\cite{monaco_water,ange_HF} and polymers \cite{fio_pb2}. More
specifically, it has been established that the decay of the
density-density correlation function occurs through different
relaxation mechanisms characterized by different time scales
\cite{scop_jpc}. These mechanisms appear as additive contributions
to the second order memory function ($M(Q,\omega)$) of the density
fluctuation spectrum. Indeed, in addition to the thermal
relaxation process, predicted by the hydrodynamic equations of a
simple fluid and arising from the coupling between density and
temperature fluctuations, one must also allow explicitly for a non
Markovian behavior of the stress tensor correlation function thus
introducing viscous relaxation. The most palpable consequence of
these relaxation processes, is that the sound propagation is no
longer determined by the hydrodynamic sound velocity.

On the basis of spectroscopic experimental evidence, viscous
relaxation is thought to proceed throughout two distinct channels
which are active on quite different time scales \cite{scop_jpc}. A
first rapid decay of $M(Q,\omega)$ (on the time-scale
$\tau_\mu(Q)$) is generally described as arising from the
interactions of each atom with its ''cage'' of surrounding
neighbors. This is followed by a slower decay process (time-scale
$\tau_\alpha(Q)$) with long lasting tails which originates from
slow, temperature-dependent structural rearrangement. Eventually,
in those systems capable of undercooling, these rearrangement
undergo a structural arrest thus giving rise to the glass
transition phenomenon. Obviously, in order to fully understand the
characteristics of these processes, which are in principle common
to many liquid systems, thorough investigations in an extended
$Q-E$ region are required. Indeed, in the recent past, massive
investigations by means of inelastic neutron scattering (INS),
devoted to liquid metals, have produced significant advances in
the comprehension of the collective properties of liquids, in
particular as far as single particle properties are concerned.
Nevertheless, with the aim of studying collective properties,
these approaches have often suffered from limitations, intrinsic
to INS, which arise mainly from the presence of incoherent
scattering and from kinematic constraints restricting the
accessible $Q-E$ region. Very recently, accurate INS experiments
have been performed \cite{cab_k,bov_k} on liquid potassium in
which the inelastic components of the spectra have been analyzed
either within a memory function framework \cite{cab_k} or by
empirical fitting functions \cite{bov_k}. In both these studies,
values of sound speed exceeding the hydrodynamic one have been
reported, and this has been interpreted as a reminiscence of
crystal-like sound propagation. This interpretation is
substantially different from the results reported in other liquid
metals (Li, Na, Al, Ga)
\cite{scop_prlli,scop_prena,scop_preal,scop_prlga} and thus
deserves a deeper investigation.

\section{the experiment}

In this work, we present an IXS study of the dynamic structure
factor $S(Q,\omega )$ in liquid potassium. Thanks to i) the purely
coherent nature of the IXS cross section in monatomic systems,
which gives direct access to the collective dynamic structure
factor, and ii) the lack of kinematic limitations, we have been
able to derive convincing information on the ultimate nature of
the atomic motion in this system. In particular, we show that, as
reported in several other simple liquids \cite{scop_jpc} for $Q$
values ranging from just below the diffraction peak maximum
($Q_{M}\approx 16$ nm$^{-1}$) all the way down to the hydrodynamic
region ($Q\approx 1$ nm$ ^{-1}$), the collective dynamics proceeds
basically through two distinct viscous relaxation processes.
Within this framework, the observed speed up of the sound
velocity, reported by the INS measurements \cite{cab_k,bov_k}, can
be interpreted as due to a relaxation mechanism. In particular,
thanks to the accessibility of the low-$Q$ and high-$E$ regions,
we show that the process mainly responsible for this increase in
the sound velocity is not related to a crossover from a
liquid-like to solid-like response. Indeed, although the system is
above the melting point, the diffusive motion is frozen on the
timescale of the density fluctuation even at the lower
investigated $Q$ values (i.~e., in terms of the Brillouin peak
frequency $\omega_B$, one finds $\omega_B(Q)\tau_\alpha(Q)>>1$
over the whole investigated $Q$ range). On the contrary, the
characteristics of the sound propagation are controlled by the
faster relaxation induced by the peculiarity of the atomic
vibrations associated with the instantaneous positional disorder
\cite{gcr_prlsim,scop_presim}.

The experiment has been carried out at the ID16 beam-line of the
ESRF in fixed exchanged wave-vector configurations. A typical
energy scan ($-50<E<50$ meV) took about 300 minutes and was
repeated several times in order to achieve a total integration
time of about 300 seconds per point. A five analyzer bench allowed
us to collect simultaneously spectra at five different values of
the exchanged wave-vector $Q$ for each single scan. The sample
thickness, ($\approx 1$ mm) was chosen in order to match the
absorption length at the incident energy of $21748$ eV,
corresponding to the (11 11 11) reflection from the silicon
analyzers. The energy resolution, $\delta E$, in the used
configuration was $\approx 1.5$ meV FWHM. The sample
was kept under vacuum at a temperature of 343$\pm$1 K without
need of any further container windows on either the beam incidence
and viewing sides.

\begin{figure}[h]
\centering
\includegraphics[width=.48\textwidth]{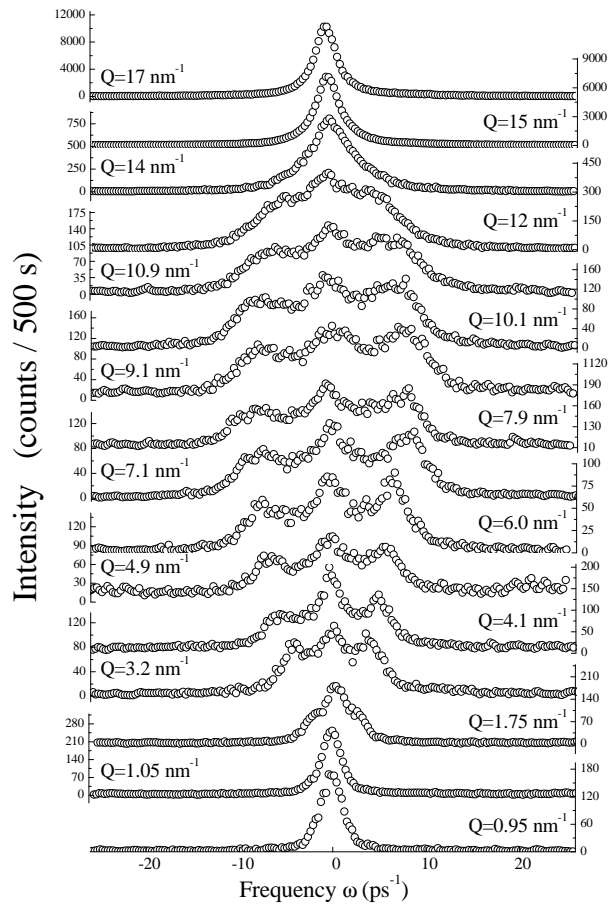} 
\caption{IXS spectra of liquid potassium ($T$=343 K) at the
indicated $Q$ values (circles)} \label{panel}
\end{figure}

In Fig.~\ref{panel} we report the measured IXS intensity for each
investigated (fixed) $Q$-value where the presence of an acoustic
propagating mode clearly appears directly from the raw data. Also
visible from the data is the presence of a maximum of the
dispersion relation (at $Q\approx$10 nm$^{-1}$) and the DeGennes
narrowing at $Q$-values close to $Q_m$, the position of the main
peak of the static structure factor $S(Q)$ ($Q_m \simeq $16
nm$^{-1}$).

\section{data analysis}

In order to extract quantitative information on these excitations,
we performed a data analysis according to the generalized
hydrodynamic \cite{BY,BALUCANI}, already tested against
experimental data available on similar systems \cite{scop_jpc}. In
this framework, the \textit{classical} dynamic structure factor
can be expressed in terms of its second order memory function,
$M(Q,t)$, so that:

\[
S(Q,\omega )=\frac{S(Q)\pi ^{-1}\omega
_{0}^{2}(Q)\tilde{M}^{\prime }(Q,\omega )}{\left[ \omega
^{2}-\omega _{0}^{2}(Q)+\omega \tilde{M}^{\prime
\prime }(Q,\omega )\right] ^{2}+\left[ \omega \tilde{M}^{\prime }(Q,\omega )%
\right] ^{2}}.
\]

where we have introduced the real ($M'(Q,w)$) and imaginary
($M''(Q,w)$) part of the Fourier transform of $M(Q,t)$, and the
frequency $\omega _{0}^{2}(Q)=K_B TQ^{2}/mS(Q)$ which is related
to the generalized isothermal sound speed $c_{t}(Q)=\omega
_{0}(Q)/Q$. This latter quantity can be calculated from the liquid
structure once $S(Q)$ is known ($m$ is the atomic mass and $K_B$
the Boltzmann constant). In order to fit the data we have
accounted for the detailed balance and for the convolution with
the instrumental resolution function $R(\omega)$ so that the
quantity $I_M(Q,\omega)$,

\[
I_{M}(Q,\omega )=\int \frac{\hbar \omega ^{\prime
}/KT}{1-e^{-\hbar \omega ^{\prime }/KT}}S(Q,\omega ^{\prime
})R(\omega -\omega ^{\prime })d\omega ^{\prime },
\]

\noindent is compared with the measured scattered intensity,
$I(Q,\omega )$.

\begin{figure}[h]
\centering
\includegraphics[width=.48\textwidth]{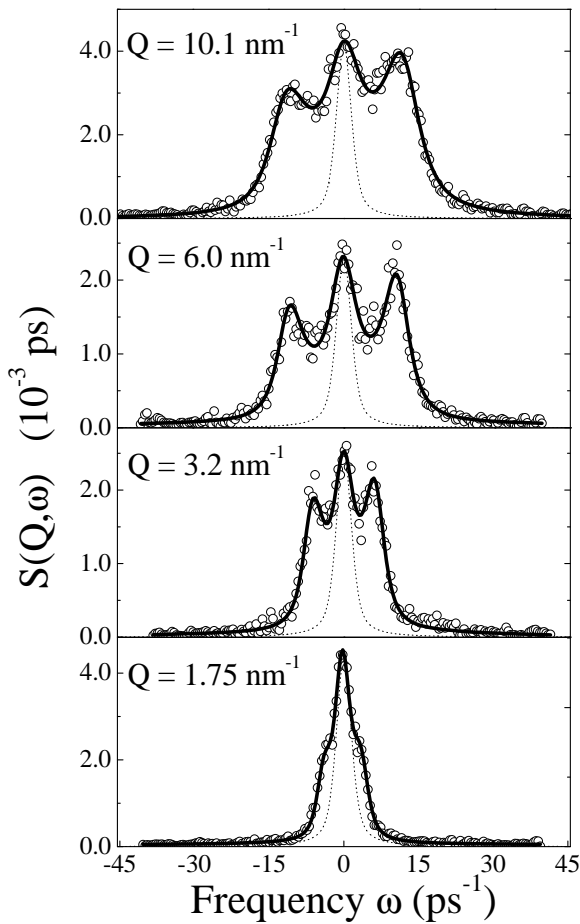} 
\caption{Experimental $S(Q,\omega)$ (open circles) for selected
$Q$ values plotted together with the fitting function (full line)
described in the text. The instrument resolution function
(full-width at half hight $\protect\delta E\approx 1.5$ meV) is
also shown (dotted line).} \label{fit}
\end{figure}

Taking advantage of the results obtained from several other liquid
metals, we utilized a memory function characterized by two
relevant time-scales associated to two processes of viscous
origin, besides the usual decay term accounting for thermal
relaxation process. This scheme is able to account for the whole
$S(Q,\omega)$ features, i.e. for both the quasielastic peak and
for the Brillouin component. To represent the two viscous terms we
adopted two simple exponential decays (corresponding to Debye
relaxations) to account for the structural ($\alpha$) relaxation
and for the faster, microscopic ($\mu$) contribution. The thermal
relaxation has been described within the usual hydrodinamics
result, which also predicts a Debye shape. Consequently, the total
memory function reads

\begin{eqnarray}
M(Q,t) &=&\left( \gamma -1\right) \omega _{0}^{2}(Q)e^{-D_{T}
Q^{2}t}
\nonumber  \label{mem} \\
&+&\Delta_\alpha ^{2}(Q)e^{-t/\tau _{\alpha }(Q)}+\Delta_\mu
^{2}(Q)e^{-t/\tau _{\mu }(Q)}.  \nonumber
\end{eqnarray}

\noindent The value of $\omega _{0}(Q)$ has been calculated using
the $S(Q)$ data reported in Ref.~\cite{OSE}, while the specific
heat ratio $\gamma$ and the thermal diffusion coefficient $D_{T}$
are derived from literature data \cite{OSE} neglecting their $Q$
dependence. This latter assumption is substantiated by the
observation that in a very similar system, namely molten lithium,
$\gamma(Q)$ (and therefore the magnitude of the thermal
relaxation), varies less then $10\%$ up to $Q_m$ \cite{can_lit}.
Although in the large $Q$ limit $\gamma(Q)$ is expected to
approach the limiting value of $5/3$, appropriate for a
non-interacting monatomic system, any recourse to
generalized-hydrodynamics arguments becomes increasingly doubtful
for wavevectors distinctly larger than $Q_m$.

All these parameters are kept fixed in the fitting procedure,
while free fitting parameters are the structural and the
microscopic relaxation times $\tau _{\alpha }$ and $\tau _{\mu }$,
and the respective relaxation strengths, $\Delta _{\alpha}(Q)$ and
$\Delta _{\mu}(Q)$. Prior to the fitting, the experimental data
have been put in absolute scale by using the first moment sum
rule.

\[
\Omega_{\tilde{S}}^{(0)}=\int\tilde{S}(Q,\omega ^{\prime })d\omega
^{\prime }=\tilde{S}(Q),
\]

\[
\Omega_{\tilde{S}}^{(1)}=\int\tilde{S}(Q,\omega ^{\prime
})\omega^{\prime} d\omega ^{\prime }=\frac{\hbar Q^{2}}{2m},
\]

The tilde indicates the dynamical structure factor corrected by
the detailed balance prefactor. In terms of the first spectral
moments of the measured intensity (subscript I) and of the
resolution function (subscript R) one can deduce that:

\begin{equation}
\tilde{S}(Q)=\frac{\hbar
Q^{2}}{2m}(\Omega_{I}^{(1)}/\Omega_{I}^{(0)}
-\Omega_{R}^{(1)}/\Omega_{R}^{(0)})^{-1}\nonumber
\end{equation}

\noindent Consequently, the normalized spectrum reads

\begin{eqnarray}
I_{N}(Q,\omega)=\tilde{S}(Q)\frac{I(Q,\omega)}{\int{I(Q,\omega)
d\omega}} \nonumber
\end{eqnarray}

A comparison between the best fitting line-shape and the
experimental spectra is reported in Fig.~\ref{fit} for selected
$Q$-values.

\begin{figure}[h]
\centering
\includegraphics[width=.48\textwidth]{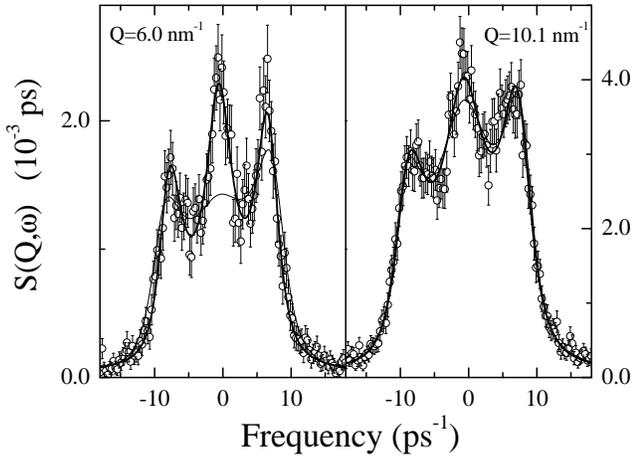} \vspace{-6.3cm}
\caption{Detail of IXS spectra of liquid potassium at Q=6.0 and
$Q=10$ $nm^{-1}$. Open circles: experimental data. Thicker line:
fit obtained by a model with two viscous relaxation channels.
Thinner line: fit results according to the usual viscoelastic
model.} \label{t1t2}
\end{figure}

It is interesting to test the applicability of the simple
viscoelastic model (single relaxation) to the experimental data.
As already noticed in several other systems, this oversimplified
approximation does not account for the lineshape details at the
level of accuracy reached by IXS (see Fig. \ref{t1t2}).

\begin{figure}[h]
\centering
\includegraphics[width=.48\textwidth]{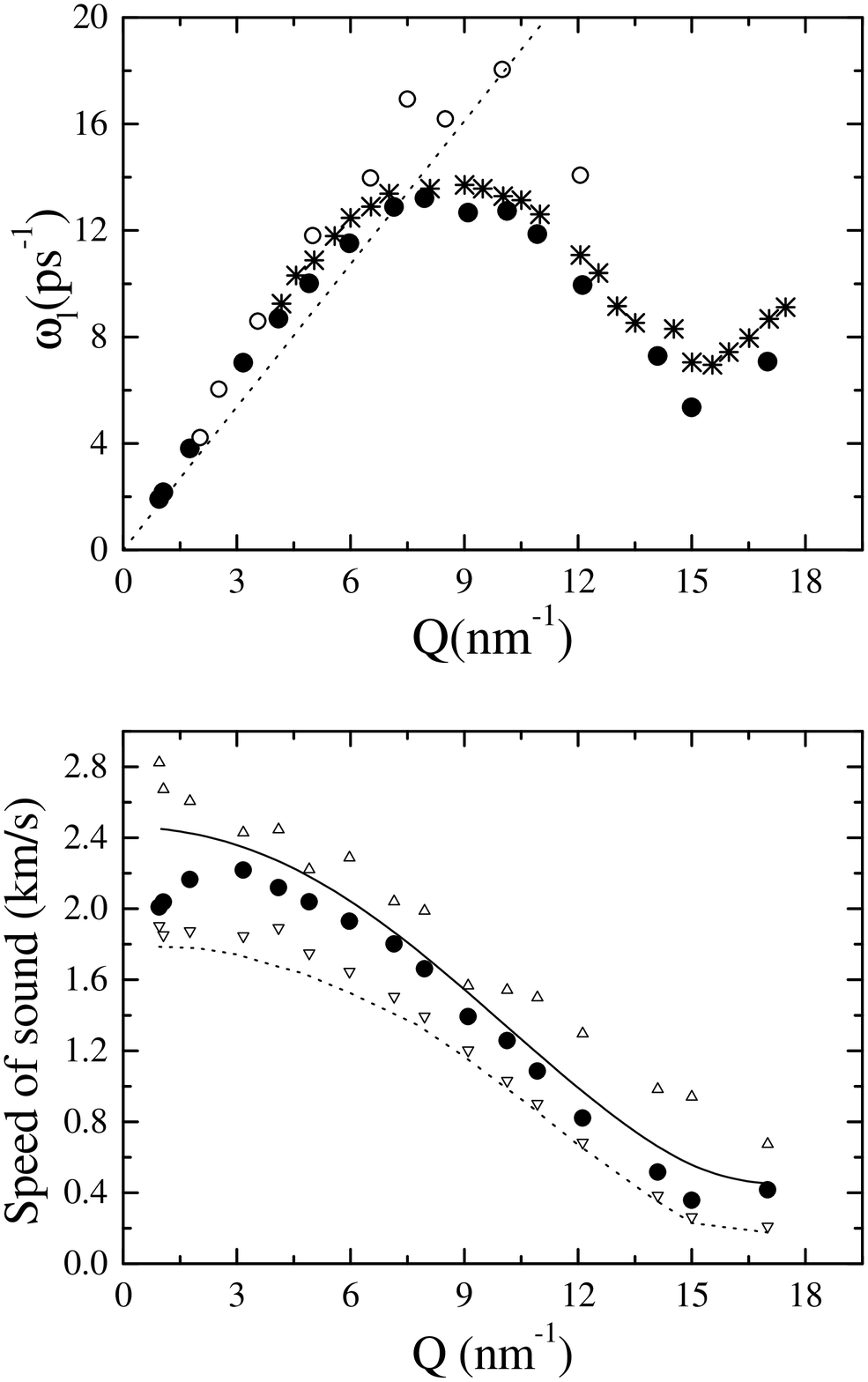}
\caption{a) The maxima of the current correlation spectra
$\Omega(Q)$ obtained from the spectra (full dots) compared with
the result of recent INS studies (stars \cite{cab_k} and open
circles \cite{bov_k}). The dotted line is the hydrodynamic
isothermal limit. b) Sound velocities deduced from the present
study: apparent ($\bullet$), $c_0(Q)$ (dotted line) and
$c_\infty(Q)$ (full line) deduced by structural data. Also
reported are the infinite frequency sound velocity $c_\infty(Q)$
($\triangle$) and the unrelaxed sound velocity of the structural
relaxation process $c_{\infty\alpha}(Q)$ ($\triangledown$) derived
from the fit.} \label{disp}
\end{figure}

One of the most relevant informations that can be obtained from an
inelastic scattering experiment is the generalized sound velocity,
defined as $c_l(Q)=\omega_l(Q)/Q$ \cite{BY,BALUCANI}. The quantity
$\omega_l(Q)$, corresponds to the position of the maximum of the
longitudinal current correlation spectrum given by $J_{L}(Q,\omega
)=(\omega ^{2}/Q^{2})S(Q,\omega )$ which we have calculated using
the classical model $S(Q,\omega )$ obtained from our best fit
procedure. In Fig.~\ref{disp}a the maxima of the current
correlation spectra (full dots) are compared with the result of
recent INS studies (stars and open circles) \cite{cab_k,bov_k}. As
can be noticed, the IXS data significantly extend at low $Q$ the
accessible kinematic region, allowing us to identify the
transition from the hydrodynamic region to the high frequency,
\textsl{mesoscopic} regime.

It is worth to emphasize that the quantity which is related to the
sound velocity is the maximum of the longitudinal current
correlation function. This is, for instance, the quantity which
continuously evolves, at increasing wavevectors, towards the single
particle regime, when distinct Brillouin modes are no longer
visible in the dynamic structure factor. Moreover, in the
framework of generalized hydrodynamics, it is the maximum of
$J_L(Q,\omega)$ which undergoes a transition between the well
defined quantities $\omega_0 (Q)$ and $\omega_\infty (Q)$
describing the relaxed and unrelaxed mechanical regimes.

In Fig.~\ref{disp}b we report the apparent sound velocity (full
dots). In the low $Q$ region, the measured apparent sound velocity
has a maximum ($c_l(Q_{max})\approx 2240$ m/s at $Q_{max} \approx
3$ nm$^{-1}$) where it clearly exceeds the isothermal value ($
\approx$ 1790 m/s) deduced from ultrasonic measurements
\cite{OSE}, a behavior (the so called positive dispersion of the
sound velocity) that is common with many other simple fluids. It
is worth to recall here that, as already pointed out in several
IXS studies (see for example Ref. \cite{scop_jpc}), in liquid
metals, owing to the very high thermal conductivity, in the $Q$
region probed here truly diffusive entropy mode are no longer
possible. In other words, the linewidth ($D_TQ^2$) of the thermal
diffusion mode exceeds the Brillouin frequency. As a consequence,
the positive dispersion of the sound velocity consists of a speed
up of the isothermal rather than the adiabatic value, as it
happens, instead, in usual liquids. In the same figure we also
report i) the generalized ($Q$-dependent) isothermal sound
velocity $\omega_{0}(Q)/Q$ (dotted line), calculated using
literature value for $S(Q)$ \cite{OSE}), that constitutes the low
frequency limit of the sound speed, and ii) the infinite frequency
sound velocity, $c_{\infty}(Q)$ (full line), related to the fourth
moment sum rule of $S(Q,\omega)$ \footnote{An accurate
determination of $c_\infty(Q)$ from the fourth moment of the
dynamic structure factor is actually prevented by the finite
instrumental resolution broadening and by background effects
which, given the finite frequency range of our data, can not be
easily accounted for} and numerically estimated using the
Price-Singwi-Tosi pseudopotential \cite{price_potalk} and a pair
distribution evaluated by molecular dynamic simulation, according
to the expression \cite{BY,BALUCANI}:

\begin{equation}
c_{\infty}(Q)=\sqrt {\frac{3K_{B}T}{m}+\frac{\rho}{mQ^{2}}\int
\frac{\partial^{2}V(r)}{\partial z^{2}} (1-e^{-iQz}) g(r) d^3 r }
\end{equation}

The up triangles are the values of $c_\infty (Q)$ deduced from the
fit as $c_{\infty}(Q)=\sqrt{\omega_{0}(Q)^2+\Delta_\alpha(Q)^2+
\Delta_\mu(Q)^2}/Q$. The down-triangles are the limiting (high
frequency) velocity associated solely to the structural relaxation
process:
$c_{\infty\alpha}(Q)=\sqrt{\omega_{0}(Q)^2+\Delta_\alpha(Q)^2}/Q$,
i.~e. the velocity which is reached when the structural relaxation
process is fully unrelaxed ($\omega_B(Q)\tau_\alpha(Q)>>1$) or, in
other words, the solid-like response. As it can be seen, the
apparent sound velocity is always larger than
$c_{\infty\alpha}(Q)$, indicating the presence of a second
relaxation process capable to drive the sound velocity from
$c_{\infty\alpha}(Q)$ to $c_{\infty}(Q)$. From this latter
observation, and from the comparison between $c_{\infty\alpha}(Q)$
and $c_{\infty}(Q)$, it appears evident how the $\alpha$ process
only plays a minor role in the full positive dispersion effect. On
the other side, it is the faster, microscopic process (for which
we find $\omega_B(Q)\tau _{\mu }(Q)\approx 1$ in the whole
examined $Q$ range), that pushes the velocity towards a fully
unrelaxed regime ($c_{\infty }(Q)$).

It is worth to mention that, in both Refs \cite{cab_k,bov_k} the
positive dispersion effect is fully ascribed to the transition
between a liquid-like and a solid-like regime, namely to the
structural relaxation process. This claim stems on the basis of
the similarity of the sound velocity value of molten potassium
with the value for the crystalline acoustic phonons along the [1 0
0] direction. This interpretation is in contrast with the results
stemming out of the present work as well as with  the analogous
findings in lithium and sodium \cite{scop_prlli,scop_prena}. In
our opinion, as the positive dispersion is a feature ubiquitous in
liquid metals, but not in the crystalline counterpart, one should
search his origin among the specific differences between the two
phases: the diffusional motion and the instantaneous disorder. As
the first candidate is ruled out by the previous considerations
(on the snap-shot timescale probed by IXS the liquid appears
frozen), it seems natural to associate this effect to the
positional disorder of the instantaneous configurations. This
experimental fact is substantiated by the evidence of positive
dispersion in the microscopic dynamics of simulated glasses
\cite{gcr_prlsim,scop_presim,kob_si02posdisp}.

Coming back to Fig.~\ref{disp}b, the slight overestimate of
$c_{\infty}(Q)$ values as obtained by the fit (up-triangles) with
respect to those derived from numerical evaluation has been
already recognized as an effect of the exponential shape of
$M(Q,t)$ in the $t \rightarrow 0$ limit \cite{fio_pb,scop_jpc}.

A further interesting issue concerns the sound attenuation
(Brillouin linewidth): once more the main actor is here the
microscopic process, which crosses the resonance condition in the
explored IXS $Q$-window ($\omega_B(Q)\tau_\mu(Q)$ always remains
of the order of one). On the contrary, the "amount of viscosity"
$\frac{\Delta^2_\alpha \tau_\alpha}{Q^2}$ associated with the
structural relaxation does not contribute to the Brillouin width,
as a consequence of the solid-like response all over the explored
$Q$ range (the Brillouin frequency always exceeds the inverse
structural relaxation time).

\begin{figure}[h]
\centering
\includegraphics[width=.48\textwidth]{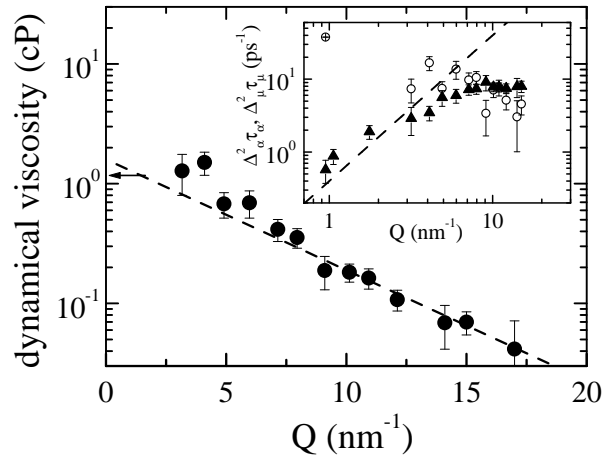}
\vspace{-6cm} \caption{a) Values of the longitudinal viscosity
$\eta_{l}(Q)$ as determined by the experimental data ($\bullet$),
dashed line is a guide to the eye. The hydrodynamic value is also
reported ($\longleftarrow$). Inset: partial contributions to the
$\eta_{l}(Q)$ due to the $\alpha$ ($\circ$)and $\mu$
($\blacktriangle$) processes, respectively. Data for the
structural process are reported for $Q\geqslant 3$ nm$^{-1}$ only,
as for lower values the determination of $\tau_\alpha$ is
unreliable due to finite resolution effects. The dashed line shows
the compatibility of the $\mu$ process with a $Q^{2}$ behavior.}
\label{visco}
\end{figure}

Within the hydrodynamic framework, the longitudinal viscosity can
be generalized accounting for its $Q-$ dependence and estimated as
the area under the total memory function. In fact, the whole
longitudinal viscosity stems from the sum of the structural and
microscopic contributions:

\begin{equation}
\eta_{l}(Q)=\rho(\Delta_{\alpha}^{2}\tau_{\alpha}
+\Delta_{\mu}^{2}\tau_{\mu})/Q^{2}.
\end{equation}

We show in Fig.~\ref{visco} as the low Q limit obtained by
extrapolation of the fitted data is in agreement with the
hydrodinamic value $\eta_l(Q\rightarrow 0)=1.11$ cP \cite{OSE}. In
the inset of Fig.~\ref{visco} we plot the single terms of the
viscosity associated with the two different processes
$\Delta_{\alpha}^{2}\tau_{\alpha}$, $\Delta_{\mu}^{2}\tau_{\mu}$;
we observe that: i) the two contributions are of the same order of
magnitude in the investigated $Q$ range, ii) the microscopic
process exhibits a Q-dependence which is compatible with a $Q^{2}$
law consistently with the data already reported in other liquids
and glasses \cite{scop_jpc}.

\section{conclusion}

We presented an experimental study of the collective high
frequency dynamics in liquid potassium at the melting temperature.
Evidence for collective acoustic modes has been found in a $Q$
region extending beyond the hydrodynamic regime up to one half of
the structure factor main peak. Far from being a mere fitting
exercise, a generalized hydrodynamic analysis allows for a
quantitative determination of relevant dynamical/thermodynamical
parameters such as the sound velocity, the structural relaxation
times and longitudinal viscosity. Although the presented approach
is based on the arbitrary choice of the functional form of the
memory kernel, it allows one to point out how many relaxation
channels are involved in the evolution of the collective dynamics
and, more important, to determine how they affect the dynamical
and transport properties.

The advantages of using a memory function based framework,
therefore, are not in a better agreement with the experimental
data but rather in the physical meaning of the involved
parameters. This approach, therefore, allows to gather information
on the physical processes underlying the relaxation dynamics, and
offers the possibility to measure physical quantities and their
finite lengthscale generalization. Through this study, for
example, we have been able to show how the Brillouin linewidth is
not related to the whole longitudinal viscosity (as it happens in
the usual light scattering regime in ordinary liquids) but rather
to the "microscopic" part of the viscosity. In other words, the
damping mechanism active in the THz region does not stem from
atomic rearrangements (the viscous flow is frozen out) but is due
to the dephasing of the density fluctuations consequent to the
non-plane wave character of the vibrational motion around the
quasi-equilibrium positions.

Along the same way, the positive dispersion effect can be regarded
as a disordered-driven relaxation process with the sound
velocity inflection occurring at the resonance condition. On the
basis of the reported findings, therefore, the interpretation of
the positive dispersion in terms of a crystal reminiscent
behavior, claimed in Refs. \cite{cab_k,bov_k} can be questioned.
The viscous to elastic transition, indeed, well observed in the
present study and in the whole alkali metal series, occurs at
frequencies below the IXS window and is quantitatively negligible
compared to the total sound velocity increase observed. In this
respect, it is worth emphasizing how, in order to gain a
quantitatively accurate determination of the sound velocity, one
has to look at the maxima of the \textit{whole} coherent current
spectra, rather than arbitrarily distinguishing between a solely
inelastic and a quasielastic contribution \cite{BALUCANI,BY}.

In conclusion, this study substantiates previous findings for the
collective dynamics in other liquid metals such as Li, Na, Al, Ga
\cite{scop_prlli,scop_prena,scop_preal,scop_prlga}. This is an
important indication of how, despite quantitative differences, the
high frequency dynamics in simple fluids exhibit universal
features which go beyond system dependent details. Since on the
observed time-scale the structure of the liquid is frozen
($\omega_B(Q)\tau_\alpha(Q)>>1$), one can think of the high
frequency dynamics as that of a system with well defined
equilibrium position (a "glass"). At the same time, the solid-like
response alone does not account for the reported spectral
features, as the details of the dynamics (sound velocity,
microscopic relaxation times, viscosity) are mainly determined by
the second, fast, relaxation process which is associated with the
vibrations of the instantaneous disordered structure, in agreement
with numerical works performed on similar systems
\cite{gcr_prlsim,scop_presim}.

\section{Acknowledgements}

Valuable help from O. Consorte and the staff of the mechanical
workshop of the University of L'Aquila (I) is kindly acknowledged.
We also thank L.E. Bove for fruitful discussions.


\end{document}